\documentclass[journal,10pt]{IEEEtran}

\IEEEoverridecommandlockouts
\usepackage{cite}
\usepackage{caption}
\usepackage[justification=centering]{caption}
\usepackage{amsmath,amssymb,amsfonts}
\usepackage{algorithmic}
\usepackage{graphicx}
\usepackage{tabularx}
\usepackage{inputenc}
\usepackage{float}
\usepackage{array}
\usepackage{makecell}
\usepackage{colortbl}
\usepackage{textcomp}
\usepackage{xcolor}
\usepackage{stfloats}
\usepackage{multirow}
\usepackage{epstopdf}
\def\BibTeX{{\rm B\kern-.05em{\sc i\kern-.025em b}\kern-.08em
    T\kern-.1667em\lower.7ex\hbox{E}\kern-.125emX}}
\begin{document}

\title{Recent Advances of 6G Ultra-Massive MIMO Technologies in Spatial and Beam Domains}

\author{Rui Feng, Cheng-Xiang Wang, \textit{Fellow, IEEE}, Jie Huang, \textit{Member, IEEE}, and Xiqi Gao, \textit{Fellow, IEEE}

\thanks{This work was supported by the National Key Research and Development Program of China under Grant 2020YFB1804901; the National Natural Science Foundation of China (NSFC) under Grants 61960206006 and 62271147; the Key Technologies R\&D Program of Jiangsu (Prospective and~Key Technologies for Industry) under Grants BE2022067, BE2022067-1, and BE2022067-2; the EU H2020 RISE TESTBED2 Project under Grant 872172; the Jiangsu Province Basic Research Project under Grant BK20192002; the Fellowship of China Postdoctoral Science Foundation under Grant 2021M690628; the Natural Science Foundation of Shandong Province under Grant ZR2019PF010; the High Level Innovation and Entrepreneurial Doctor Introduction Program in Jiangsu under Grant JSSCBS20210082; and the Fundamental Research Funds for the Central Universities under Grant 2242022R10067.}

\thanks{R. Feng is with Purple Mountain Laboratories, Nanjing, 211111, China, and also with the School of Information Science and Engineering, Southeast University, Nanjing, 210096, China (e-mail: fengxiurui604@163.com).}

\thanks{C.-X. Wang (corresponding author), J. Huang, and X. Q. Gao are with the National Mobile Communications Research Laboratory, School of Information Science and Engineering, Southeast University, Nanjing, 210096, China, and also with the Purple Mountain Laboratories, Nanjing, 211111, China (e-mail: $\left\{\text{chxwang, j\_huang, xqgao}\right\}$@seu.edu.cn).}

}

\markboth{IEEE Vehicular Technology Magazine,~Vol.~XX, No.~XX, MONTH~2022}%
{R. Feng \MakeLowercase{\textit{et al.}}: Bare Demo of IEEEtran.cls for IEEE Journals}

\maketitle

\begin{abstract}
To explore the full potential of ultra-massive multiple-input multiple-output (MIMO) communication systems, it is fundamental to understand new ultra-massive MIMO channel characteristics and establish pervasive channel models.
On this basis, large dimensional spatial-temporal transmission and random access technologies need to be investigated and evaluated for better practical implementation.
Firstly, this paper reviews recent advances of ultra-massive MIMO technologies in the traditional spatial domain, including wireless channel characterization and modeling, channel estimation, spatial multiplexing, and precoding.
Secondly, considering the dramatic increase of base station (BS) antennas and access users in ultra-massive MIMO systems, the confronted high dimensional complexity and computing burden of these ultra-massive MIMO technologies are indicated.
To provide efficient and systematic solution, the emerging tendency to transform related technologies from the traditional spatial domain to beam domain is introduced. The utilities of large sparsity merit, reduced energy consumption, and improved usage of radio frequency (RF) chains in the beam domain channel are elaborated.
At last, future challenges of ultra-massive MIMO communication systems are discussed.
\end{abstract}

\begin{IEEEkeywords}
6G, ultra-massive MIMO, spatial domain, beam domain, physical layer technologies.

\end{IEEEkeywords}

\section{Introduction}

To realize the visions of the sixth generation (6G) wireless communication system, full utilization of spatial, temporal, frequency, and code resources have been explored to serve ultra-massive transmission and access of users/devices and to provide high quality communication even in very high mobility scenarios \cite{J2020_SCIS_XHYou_6G, J2020_VTM_WangCX}.
Ultra-massive multiple-input multiple-output (MIMO), evolving from long time evolution (LTE) MIMO and massive MIMO, equips hundreds or thousands of antenna elements at base station (BS) can serve tens or hundreds of users with the same time and frequency resources simultaneously. Benefit from the abundant spatial degree of freedom, the ultra-massive MIMO technology is a perfect candidate catering to 6G demands. It can greatly increase channel capacity, energy efficiency, and spectrum efficiency.
However, it also poses challenges to hardware design, signal and arithmetic design and processing, resource allocations, etc.

\begin{figure*}[tb]
	\centerline{\includegraphics[width=0.9\textwidth]{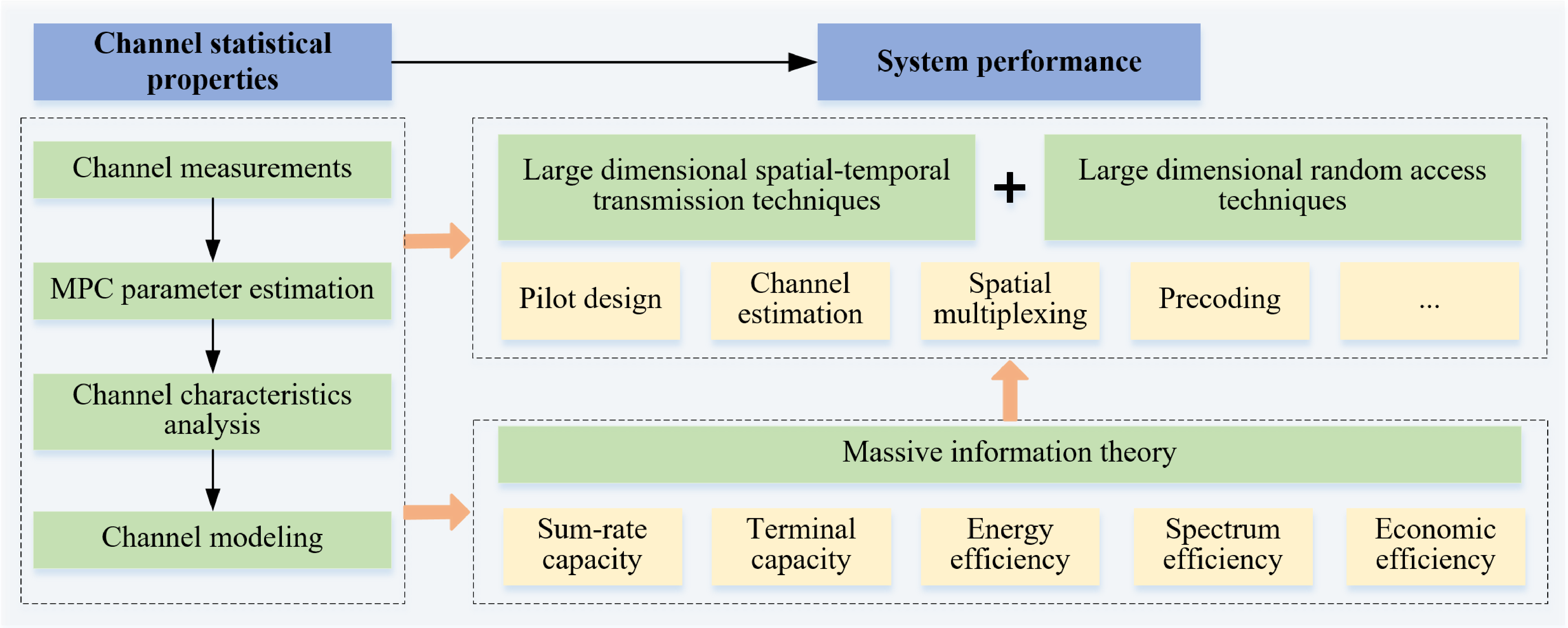}}
	\caption{Key PHY technologies of ultra-massive MIMO.}
	\label{Fig_PHY_Theory}
\end{figure*}

The design of ultra-massive MIMO system is highly relying on fundamental physical layer (PHY) theories and technologies.
As shown in Fig.~\ref{Fig_PHY_Theory}, wireless channel measurements, multipath component (MPC) parameter estimation, channel characteristics analysis, and channel modeling are essential procedures to understand the statistical properties of ultra-massive MIMO channel. They are useful to clarify the dependency relationships of channel statistical properties with array configurations, frequency bands, and other system setups. A pervasive channel model is also an important pre-requisite for the following PHY technology design and evaluation.
Based on the channel characteristics and channel model, large dimensional spatial-temporal transmission techniques and random access techniques can be studied, including channel estimation, spatial multiplexing, precoding, etc \cite{J2017_TAP_JBrady}. In addition, the impacts of system configurations on energy efficiency, spectrum efficiency, economic/hardware efficiency, sum-rate capacity, and terminal capacity can be analyzed with the aid of massive information theory.

However, with the ultra-massive expansion of array size, traditional methods and techniques in MIMO and even massive MIMO systems are no longer suitable for ultra-massive MIMO system.
It has already been verified that wireless channels facilitating ultra-massive array show different propagation characteristics, including more obvious spherical wavefront, spatial non-stationarity, channel hardening, etc.
Consequently, traditional MPC parameter estimation algorithms and channel models under plane wavefront and spatial stationary assumptions are not suitable for ultra-massive MIMO channel.
In addition, as the large amount of antenna elements require paramount radio frequency (RF) chains, this poses heavy burden on practical implementation as well as channel state information (CSI) acquisition.
Due to the imperfect channel estimation under massive pilot overhead, the upper bound of massive spatial-temporal transmission sum-rate and the tradeoff among energy, spectrum, and economic efficiencies need to be studied.
Large dimensional spatial-temporal transmission and random access techniques need to be explored to adapt to more complex communication scenarios.
It is also necessary to work on large dimensional computing theory to support reliable communication in real time.
Then, how to reduce complexity with ensured accuracy at each key procedure is crucial for ultra-massive MIMO system.

Traditionally, it is natural to study wireless channel and PHY techniques in the spatial domain.
Recently, more efforts have been devoted to researches in the beam domain.
The beam domain channel can be transformed from traditional spatial domain channel using unitary discrete Fourier transformation (DFT) matrices.
The wireless channel can be observed through transmitter (Tx) and receiver (Rx) beam pairs.
The channel coefficient of each beam pair is only contributed by MPCs that fall into this beam pair, i.e., with limited angles of arrival and departure.
Considering the clustering nature of MPCs in ultra-massive MIMO channels, only a few channel coefficients are related to MPCs, while others are dominated with noise. This is also known as the channel sparsity.
It is found that the different beam domain channel elements are asymptotically uncorrelated and their envelops tend to be independent of time and frequency \cite{J2017_JSAC_LYou}.
To explore the full potential of ultra-massive MIMO communication systems, it is meaningful to fully review the recent advances of ultra-massive MIMO channel characterization and modeling, as well as large dimensional spatial-temporal transmission and random access techniques in both spatial and beam domains.

The remainder of this paper is organized as follows.
In Section~\uppercase\expandafter{\romannumeral2}, we focus on the ultra-massive MIMO channel characterization and modeling.
In Section~\uppercase\expandafter{\romannumeral3},  ultra-massive MIMO PHY technologies are surveyed in both spatial and beam domains.
In Section~\uppercase\expandafter{\romannumeral4}, future research challenges are given.
Finally, conclusions are drawn in Section~\uppercase\expandafter{\romannumeral5}.

\section{Ultra-massive MIMO Channel Characterization and Modeling}

\subsection{Channel characterization and modeling in the spatial domain}

The comprehensive understanding of ultra-massive MIMO channel characteristics and acquisition of wireless channel mathematical expression are meaningful for the design of devices, antennas, radio and signal processing elements, protocols, and systems/networks according to system requirements.
It is fundamental to evaluate link-level, system-level, and network-level performance in a realistic, repeatable, and reproducible manner \cite{ZLv22_TVT}.

The channel characterization and modeling rely on carrying out extensive channel measurements, high resolution parameter estimation, and channel characteristics analysis.
Especially for MPC parameter estimation, besides the Bartlett beamforming, Capon's beamformer, multiple signal classification (MUSIC) algorithm, unitary estimating signal parameter via rotational invariance technique (ESPRIT), etc., the commonly used algorithm is the space-alternating generalized expectation-maximization (SAGE) algorithm.
However, with the expansion of antenna elements and high temporal resolution brought by millimeter wave (mmWave) and terahertz (THz) bands, the data size and number of parameters to be estimated increase exponentially. This poses huge challenge to real time data processing, especially in high mobility time-variant scenario.
In addition, some new ultra-massive MIMO channel characteristics need to be considered.
For example, the Rayleigh distance calculated under ultra-massive array becomes large, and cluster/receiver lying within the Rayleigh distance suffers from the near field effect. That means, the impinging wave exhibits spherical wavefront and arrives at ultra-massive antenna elements with different angles and delays. Besides, different antenna elements may see different clusters, i.e., there is spatial non-stationarity along the ultra-massive array \cite{J2020_VTM_WangCX}.
While most existing channel parameter estimation algorithms assuming narrowband, plane wavefront, and spatial stationary propagation, researches on algorithms that considering new massive MIMO channel characteristics need to be considered.

Wireless channel models can be divided into traditional non-predictive and artificial intelligence (AI) based predictive categories \cite{RFeng_IoT22}.
The former category includes deterministic model, correlation based stochastic model (CBSM), and geometry based stochastic model (GBSM).
Deterministic model and CBSM perform less satisfactory in complexity and accuracy, respectively.
GBSM is widely used to provide more flexible and accurate channel modeling.
It has already been used to describe new channel characteristics of (ultra-)massive MIMO channels. For example, the near field spherical wavefront and spatial non-stationarity have been modeled by deriving the MPC travel distance with angle of arrival/departure (AoA/AoD), and visible region/cluster birth-death process, respectively~\cite{ZLv22_TVT}.
Likewise, with the increase of antenna elements, the complexity of GBSM becomes not affordable.
Benefitting from the non-linear big data processing advantages, the latter AI based predictive methods can not only be used to generalize channel characteristic parameters, but also predict channel characteristics at unknown frequency bands and future times. The used algorithms include multi-layer perceptron (MLP) neural network (NN), convolutional NN (CNN), recurrent NN (RNN), long short-term memory (LSTM), random forest, etc. However, the accuracy of AI based predictive methods is still worth further investigating.

\subsection{Channel characterization and modeling in the beam domain}

\begin{figure}[t]
	\centerline{\includegraphics[width=0.48\textwidth]{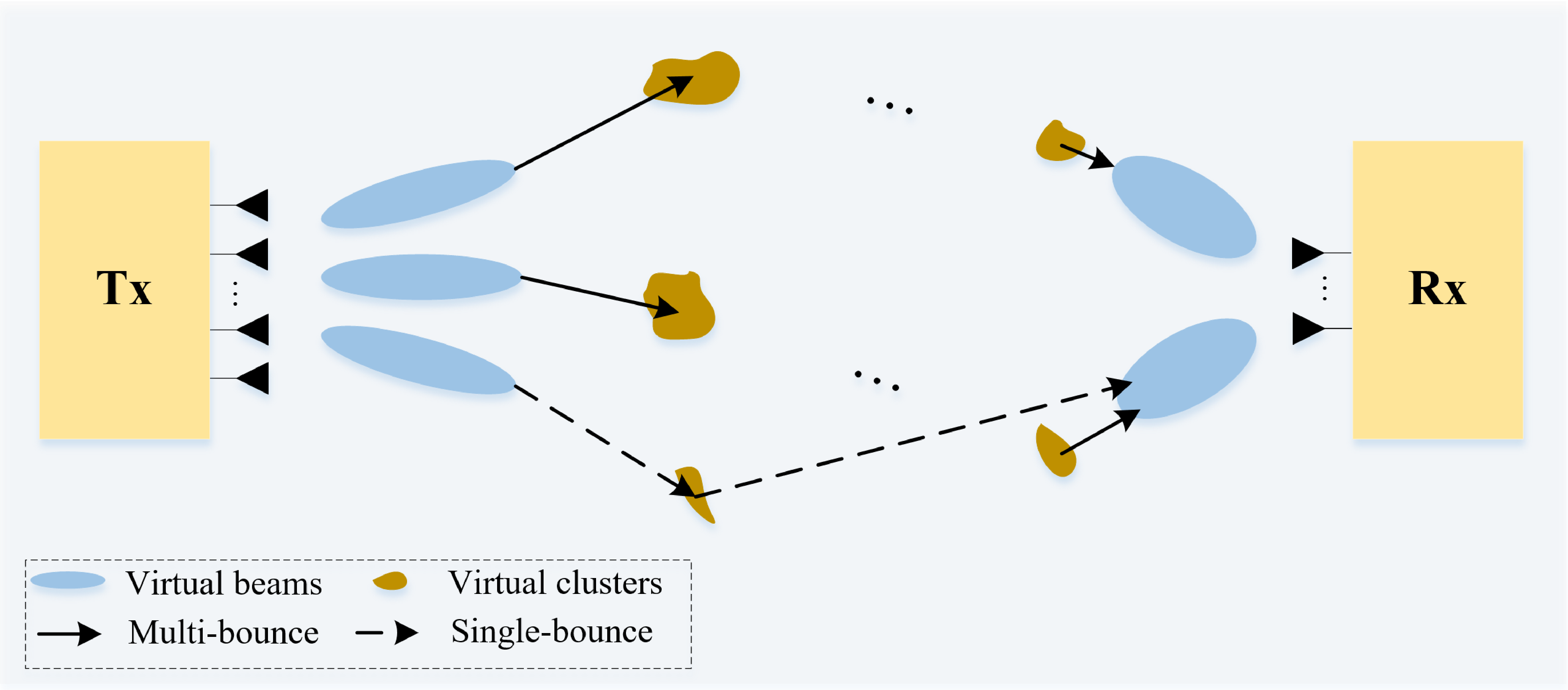}}
	\caption{Wireless channel in the beam domain.}
	\label{Fig_BeamDomain_Channel}
\end{figure}

As shown in Fig.~\ref{Fig_BeamDomain_Channel}, beam domain channel is formed by dividing the whole space into several virtual sections according to the three-dimensional (3D) angles.
The corresponding cluster is virtual, which contributes to the transmission of signals from one given direction to another.
The beam domain channel can also be transformed from the traditional spatial domain by using two unitary DFT matrices.

There are some researches that performing AoA estimation in the beam domain. The proposed algorithms mainly include beam domain MUSIC, ESPRIT, unitary ESPRIT, SAGE algorithm, etc.
They formulate the beam domain signal model by multiplying DFT matrices to that in the spatial domain.
Especially in massive MIMO channels, the necessary high dimensional matrix operations can be alleviated to magnitudes of beam number, rather than antenna number. Therefore, the complexity of channel parameter estimation can be greatly reduced.
It can further reduce the computational complexity by considering the limited angular spread and confining the beams to interested sectors.
In \cite{J1996_TSP_MDZoltowski}, a beam domain unitary ESPRIT algorithm was introduced by using DFT matrix to transform data into real-valued beam domain. It indicated the benefits of beam domain channel parameter estimation: reduced computational complexity, less sensitivity to antenna array imperfection, and lower signal-to-noise ratio (SNR) resolution threshold.
The utilization of beam domain parameter estimation is of more value in ultra-massive MIMO since the antenna number is quite large.
However, existing literature lacks multi-dimensional parameter estimation and needs to be extended for ultra-massive MIMO channels.

\begin{figure}[]
	\centerline{\includegraphics[width=0.5\textwidth]{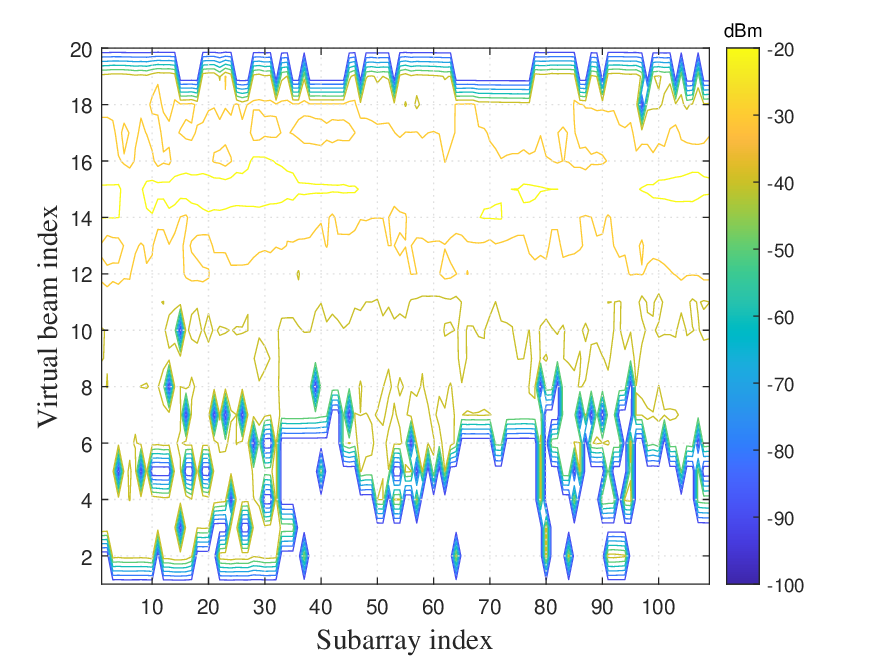}}
	\caption{Power variations of different virtual beams.}
	\label{Fig_PowerVariances}
\end{figure}

\begin{figure*}[]
	\centerline{\includegraphics[width=0.8\textwidth]{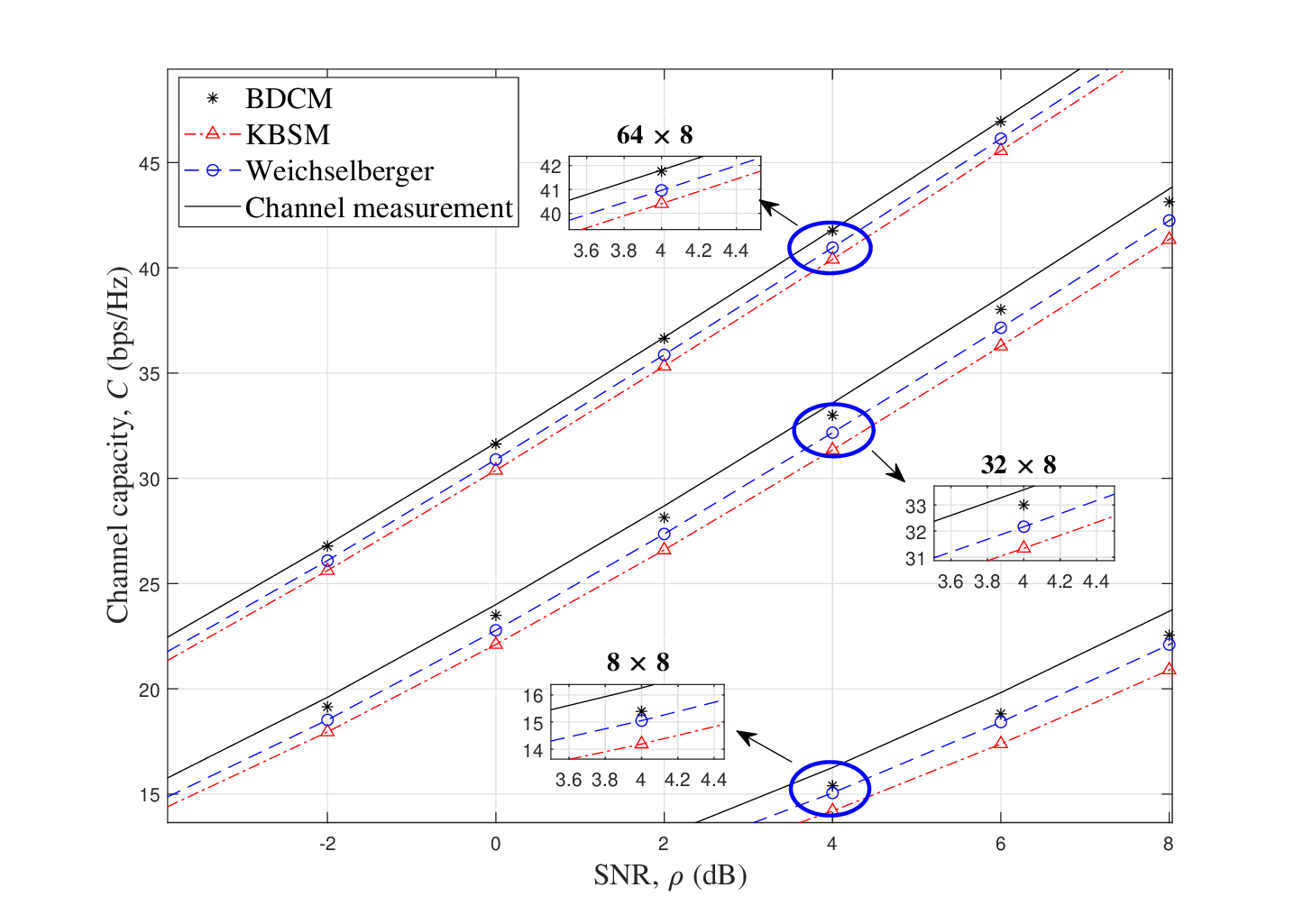}}
	\caption{Channel capacities of various channel models.}
	\label{Fig_Capacities}
\end{figure*}

For the convenience of beam domain channel modeling, ultra-massive MIMO channel characteristics in the beam domain need to be studied.
To explore the differences of channel characteristics in the beam domain with those in traditional array domain, we have conducted ultra-massive array channel measurements and data analysis.
The Rx with 128 elements uniform linear array (ULA) is elevated to the outside wall of a building with 20 m height and the Tx equipped with 8 antennas is positioned on the truck with 1.5~m height. The carrier frequency is 5.3 GHz and the bandwidth is 160 MHz.
As shown in Fig. \ref{Fig_PowerVariances}, by employing a sliding window with size 20 and step 1, the virtual beam powers at different subarrays are illustrated in a contour map. It can be seen that the dominant MPCs fall within beams 12--18. The powers of these beams are not shifted regularly to other adjacent beams. Therefore, near field spherical wavefront is not significant in beam domain. However, other new beam domain channel characteristics need to be studied based on extensive channel measurements and data processing.

Beam domain channel model (BDCM) is originated from the virtual channel representation (VCR), which was proposed in 2000 by Akbar M. Sayeed \cite{J2002_TSP_AMSayeed}. BDCM characterizes MIMO channels by partitioning the propagation channels into virtual beamspaces. Recently, this model has been developed for massive MIMO channels. In this work, we unify the definition of BDCM and VCR as BDCM.
An usual way to establish BDCM of massive MIMO channel is firstly set up a GBSM and then transform it into beam domain by using two unitary matrices.
However, when considering spherical wavefront or spatial non-stationarity, it is not easy to derive the corresponding unitary matrices.
Another way is to sum MPC contributions into certain beams up and to analyze beam domain signal propagation mechanisms, namely establish the BDCM directively based on beam domain channel characteristics.
Preliminary work on this method can be found in the literature, though new ultra-massive MIMO channel characteristics have not been considered.
In Fig. \ref{Fig_Capacities}, channel capacities of the Kronecker based stochastic model (KBSM), Weichselberger model, and BDCM are compared with that of the channel measurement data. Three Rx array sizes are employed, i.e., 8, 32, and 64.
Note that the BDCM is transformed from a 3D massive MIMO GBSM through Fourier matrices. It can be seen that BDCM performs better than the Weichselberger model and KBSM, especially when the Rx array size increases. This is because that, with the increased antenna number, the virtual beams of the BDCM become very narrow and the BDCM performs approximately with the GBSM.
It needs to be noted that, here we have not taken the new ultra-massive MIMO characteristics into consideration, which can further improve the BDCM performance.

\section{Ultra-massive MIMO PHY technologies}

\begin{figure*}[tb]
	\centerline{\includegraphics[width=1\textwidth]{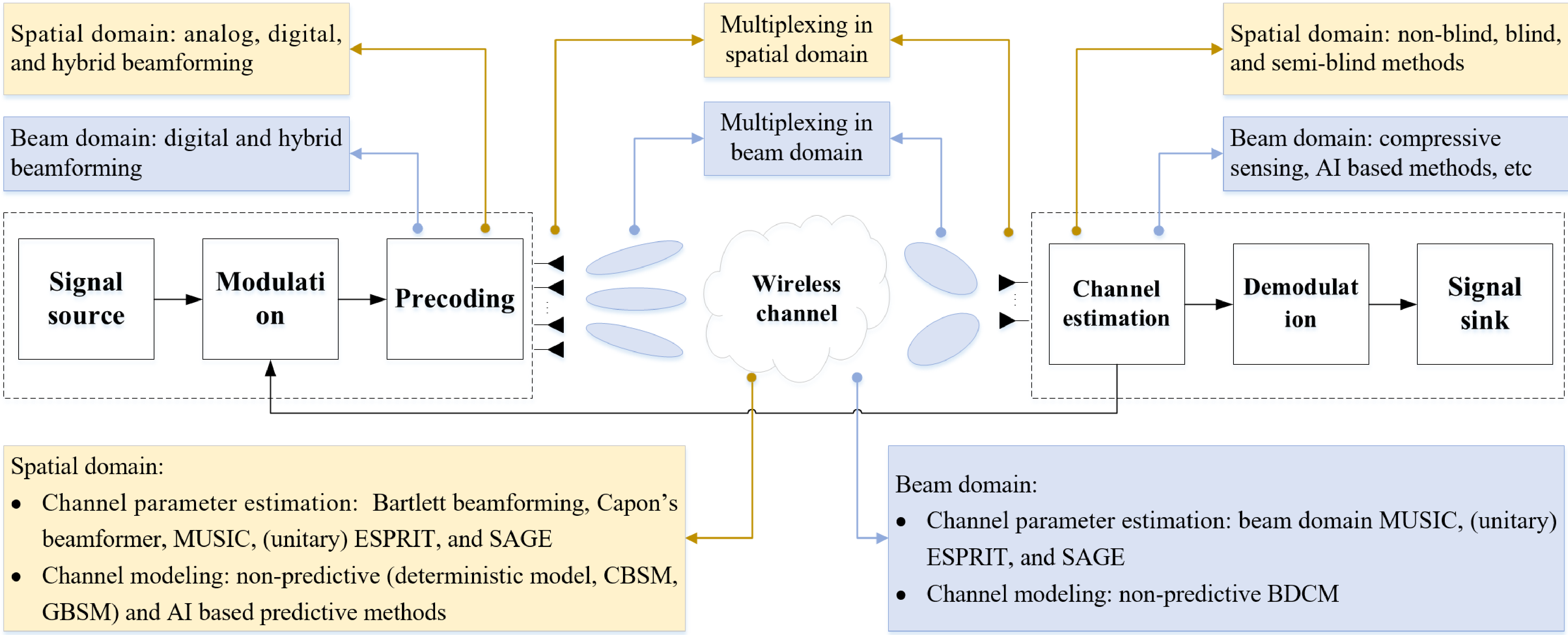}}
	\caption{Working flowchart of ultra-massive MIMO.}
	\label{Fig_MMIMO_Technol}
\end{figure*}

In Fig.~\ref{Fig_MMIMO_Technol}, we illustrate the working flowchart of ultra-massive MIMO system.
After careful signal design, signals are firstly modulated and precoded according to the estimated CSI. With the inserted cyclic prefix, signals are then transmitted into the wireless channel and received by antenna elements. Then, the cyclic prefix is removed and equalization can be performed before demodulation and reception.
As techniques related to wireless signals and channels can be performed in both spatial and beam domains, in the following, we will focus on giving a review of channel estimation, spatial multiplexing, and precoding techniques.

\subsection{Channel estimation}

\subsubsection{Channel estimation in the spatial domain}

Perfect CSI is crucial for the BS to steer beams to desired directions.
With perfect CSI estimation, the transmission power efficiency can be proportional to the BS antenna number, vice versa for the downlink.
To obtain the CSI under time division duplex (TDD), the BS needs to send pilot signals to the Rx, and the receiver performs channel estimation and sends feedback.
The uncorrelated interference and pilot contamination are main factors that can impact CSI estimation performance.
Accurate channel acquisition relies on efficient channel estimation algorithms.
There are blind estimation and semi-blind estimation. The former requires the signal covariance matrix rather than the dedicated pilot signal.
Minimum mean square error (MMSE) algorithm can be used to provide power efficiency that is proportional to the square root of the BS antenna number.
Channel estimation can also be operated based on the estimated AoA. This method avoids pilot contamination, thus improves spectrum efficiency. However, this method needs extra array calibration and AoA estimation.
Considering that the acquisition of instantaneous CSI is a huge task, statistical CSI can be considered instead when the second order statistics of CSI varying slowly.

However, the utilization of excess antennas at BS of ultra-massive MIMO communications brings overwhelming pilot overhead and computational complexity, even for statistical CSI \cite{J2020_WC_GaoXQ}. The high dimensional signals and parameters to be estimated hinders the boosting of transmission efficiency.
Researchers also resort to AI algorithms to provide accurate estimation with limited samples.
Specifically, in fast moving environments, e.g., vehicle-to-vehicle, high speed train, and unmanned aerial vehicle, wireless channels change frequently and coherence times are very small.
In order to provide accurate CSI for the BS, frequent channel estimation needs to be performed.
However, the pilot overhead is overwhelming and the complexity is magnified especially with the increased antenna number in ultra-massive MIMO channel.
To tackle this problem, the sparsity of ultra-massive MIMO channel and the lower variants nature of AoA/AoD than path gains can be utilized to estimate time-varying channel.
There are also compressive sensing and machine learning based algorithms that can be used for time-varying ultra-massive channel estimation, e.g., message passing, Gaussian-mixture Bayesian learning, etc.
Channel estimation under frequency division duplex also needs to be investigated.

\subsubsection{Channel estimation in the beam domain}

Utilizing ultra-massive MIMO is challenging in terms of the unaffordable hardware complexity and energy consumption.
There are many works that employing the lens array to transform the conventional antenna domain channel into beam
domain.
Thus, the number of RF chains can be reduced.
Resorting to the sparsity of ultra-massive MIMO channel, especially dominant in mmWave and THz bands, it was introduced in \cite{J2020_Network_WangN} that the sparsity of virtual beams can be used to detect pilot contamination attacks and improve channel estimation results. The beam domain channel estimation can also alleviate the constraints on RF chains.

However, considering the large lens antenna array and reduced RF chains, channel estimation confronts extremely challenges.
Compressive sensing method has been used to estimate channel by utilizing the channel sparsity in beam domain.
For example, orthogonal matching pursuit and compressive sampling matching pursuit algorithms have been used in digital beamforming, hybrid beamforming, etc.
In \cite{J2021_TC_DaiLL}, a support beam detection based beam domain channel estimation scheme was proposed.
It considered the sparse virtual beam channels individually. Through the support beam detection, different pilot overheads can be assigned for different virtual beam channels.
Therefore, superior performance can be obtained especially in low SNR condition.
To utilize the sparsity and MPC clustering natures of ultra-massive MIMO channels, algorithms for image reconstruction have also been introduced.
Further, there are also AI based beam domain channel estimation methods for massive MIMO channel to provide better accuracy. For example, there are approximate message passing (AMP) based method, Gaussian mixture learned AMP method, and denoising CNN.

\subsection{Multiplexing}

\subsubsection{Multiplexing in the spatial domain}

In MIMO systems, channel capacity and spectrum efficiency can be ensured by the spatial multiplexing technique \cite{J2020_WC_ShenS}.
By grouping outgoing signal into multiple streams, multiplexing gain can be obtained by transmitting multiple data streams over parallelled sub-channels.
At the Rx side, antenna number needs to be no less than the multiple data stream number, thus signals can be correctly decoded and channel capacity can be improved.
The high-rate spatial multiplexing is also beneficial for the following spatial modulation and beamforming.

In ultra-massive MIMO systems, unprecedented spatial multiplexing capabilities are expected considering the high spatial resolution and favorable propagation condition.
However, the numbers of RF chains at Tx and Rx are usually limited due to hardware complexity.
Then, the multiplexing gain is restricted by the smaller number of RF chains.
In addition, spatial multiplexing relies on the independence among array elements or sub-channels, it handles uncontrolled MPC signals.
Whereas with limited scattering, the ultra-massive MIMO channels are highly correlated and MPC channel matrices are low-rank.
In order to boost the multiplexing gain, new mechanisms that are tailored for ultra-massive MIMO channels needs to be considered.

\subsubsection{Multiplexing in the beam domain}

Beam domain multiplexing can leverage the benefits of beamforming and multiplexing, thus to obtain improved array gain and multiplexing gain.
In \cite{J2020_WC_ShenS}, it was indicated that beam domain signal processing, especially the beam domain multiplexing, will play a vital role in future wireless communication system.
It was said that beam domain multiplexing can be used to improve channel capacity and provide better coverage for cell-edge users.
The authors proposed a beam domain multiplexing technique and compared that with the spatial domain multiplexing in detail.
It was shown that beam domain multiplexing relies on the independence among multiple virtual beams and the correlation of the single beam.
Beneficial from the beamforming, it handles multiple dedicated directional signals rather than uncontrolled multipath signals.

In ultra-massive MIMO system, the utilization of beam domain multiplexing has not been extensively studied.
Taking the new channel characteristics of ultra-massive MIMO channel into consideration, it needs to be noted that the interference-rejecting and dynamic beamforming techniques are required.

\subsection{Precoding/beamforming}

Traditional precoding/beamforming techniques rely on the accurate acquisition of CSI.
With known CSI, precoding is operated at the Tx side to adjust data streams for better adaption to the channel. It can be used to focus radiation energy into target direction, thus to improve energy efficiency and decrease multi-user interference. Before transmission, precoding can be performed by post-multiplying the channel matrix with a precoding matrix. The precoding matrix varies with different methods.
There are linearly zero forcing (ZF), block diagonalization, and matched filter (MF) methods, and non-linear dirty paper coding (DPC), vector perturbation, and lattice-aided methods \cite{J2013_SPM_Rusek}.
ZF is the simplest method and when the BS antenna number grows, it performs as well as the interference-free system. However, the computational complexity to get the matrix pseudoinverse is high.
MF can achieve high signal to interference plus noise ratio when BS antenna number scaling up.

As analog beamforming utilizes analog phase shifters to acquire beamforming gain, it is not suitable for multiple data stream transmission.
By using digital processing chain for each antenna element, full digital beamforming can deliver high energy efficiency, channel capacity, and relaxed hardware requirements. However, with mmWave and THz bands, the wide bandwidth requirement on digital-to-analog and analog-to-digital converters hinders the usage of this method.
Hybrid beamforming, namely analog-digital beamforming, has been proposed to provide low implementation complexity and reduced power consumption. It also eliminates CSI estimation.
It consists of analog multi-beam selection and blind estimation algorithm.
It was indicated in \cite{J2017_TAP_JBrady} that the utilization of beam domain MIMO technologies with hybrid precoding, near-optimal performance with dramatically reduced complexity can be achieved.

In ultra-massive MIMO channels, the sparsity property is expected to approximate full digital beamforming performance.
There are also AoA based precoding/beamforming techniques that do not need CSI. However, the AoA estimation needs to be performed for data transmission.
To settle the interference problem of closed users in signal separation, constant modulus algorithm, independent component analysis, etc., can be used.
Distributed MIMO can efficiently exploit the diversity and offer excellent coverage. Cell-free ultra-massive MIMO, facilitating with large amount of distributive access points to serve small users, inherits advantages from both distributed MIMO and ultra-massive MIMO. Considering the reduced computational complexity and the distributed implementation merit, conjugate beamforming is very useful for signal processing at distributed access points. It is an important candidate for cell-free ultra-massive MIMO network. However, the increased self-interference in forward-link transmission needs to be solved by either assigning pilots for each user or introducing pilot recovery technique.

\subsection{Comparison of PHY technologies in spatial and beam domains}

Resorting to the different channel characteristics in spatial and beam domains, principles of transmission technologies show some differences.
The privileges of ultra-massive MIMO technologies in beam domain over spatial domain are briefly illustrated in Table \ref{Table_Comparison}.
The overall concerns of spatial domain data processing are the complexity and computing burden brought by the increased antenna array number.
Besides providing reduced complexity, the execution of beam domain processing can also provide concise expression and alleviate hardware constrains, for example in channel estimation.

\begin{table*}[]
\caption{ADVANTAGES AND CHALLENGES OF ULTRA-MASSIVE MIMO TECHNOLOGIES IN THE BEAM DOMAIN.}
\label{Table_Comparison}
\renewcommand{\arraystretch}{1.5}
\begin{tabular}{|m{3cm}<{\centering}|m{6.2cm}<{\centering}|m{6cm}<{\centering}|}
\hline

\textbf{Technology}                & \textbf{Advantage}      & \textbf{Challenge}   \\   \hline
\textbf{MPC parameter estimation}       & \makecell[l]{$\bullet$ Reduced complexity \\ $\bullet$ Less sensitivity to array imperfection \\ $\bullet$ Decreased SNR resolution threshold}     & \makecell[l]{$\bullet$ Full exploration of channel characteristic \\ $\bullet$ Multi-dimensional parameter estimation}      \\  \hline
\textbf{Channel modeling}                    & \makecell[l]{$\bullet$ Reduced complexity \\ $\bullet$ Linearly relationship with angle \\ $\bullet$ Concise expression for system performance analysis}   & \makecell[l]{$\bullet$ Consideration of new channel characteristic \\ $\bullet$ Tradeoff between accuracy and complexity}        \\  \hline
\textbf{Channel estimation}                   & \makecell[l]{$\bullet$ Reduced pilot overhead \\ $\bullet$ Decreased computational complexity  \\ $\bullet$ Alleviated RF chain constraints}   &  \makecell[l]{$\bullet$ Accuracy with limited RF chains \\ $\bullet$ Exploration of efficient algorithm} \\ \hline
\textbf{Spatial multiplexing}                 &  \makecell[l]{$\bullet$ Improved array gain \\ $\bullet$ Improved multiplexing gain}  &   \makecell[l]{$\bullet$ Necessities of interference-rejecting and dynamic \\beamforming techniques}     \\  \hline

\end{tabular}
\centering
\end{table*}

\section{Future Research Challenges}

\subsection{Ultra-massive antenna array configurations}

Ultra-massive linear array could stretch tens or hundreds meters long at sub-6 GHz frequency band. In practical implementation, this poses serious requirements on related infrastructures such as the antenna derrick. Besides, ULA can only offer azimuthal spatial resolution, while unable to provide elevational coverage.
To relax such pressures, more compact 3D array configurations can be used, such as planar array, circular array, and cylindrical array. However, mutual coupling effect that may have impact on beamforming needs to be mitigated.
The utilization of mmWave and THz can also facilitate ultra-massive antennas into a more compact size \cite{J2017_TWC_Ngo}.
Besides the aforementioned centralized antenna array, there are also distributed antenna array. The antenna elements can be separately placed to improve signal coverage.
As a special case of distributed array, block array is introduced to provide more flexible for configuration.
To avoid the intercell interference, cell-free massive MIMO was proposed.
In cell-free network, many access points controlled by a central processing unit are distributed coherently to serve many users with the same time and frequency resources \cite{J2021_CM_Matthaiou}. It is foreseen as a promising network to clarify the boundary effect of conventional cellular networks and to provide huge energy and spectrum efficiencies with simple signal process technique.

\subsection{Holographic massive MIMO and reconfigurable intelligent surface (RIS)}

Considering that channel capacity can grow monotonically with the increase of antenna number, holographic massive MIMO has been proposed to approach the upper bound limits of massive MIMO system.
RIS is a meta-surface that can flexibly steer the signal transmission direction and improve signal coverage \cite{J2021_CM_Matthaiou}.
In recent years, there are more works using massive array in RIS. It has been proposed that by facilitating enough antennas at the RIS, a continuous surface can be formed to be holographic RIS. Though this is an ideal case, the utilization of massive array is destined.
With the intervention of RIS in ultra-massive MIMO channel, related PHY technologies need to be changed correspondingly, including channel modeling, channel estimation, etc.
In general, holographic massive MIMO and RIS are exciting and challenging research topics.

\subsection{Beam division in temporal and frequency domains}

It has been pointed out in \cite{J2002_TSP_AMSayeed} that, except dividing virtual beams in spatial domain, we can also divide beams in both temporal and frequency domains.
Therefore, by using the joint virtual beams in spatial-temporal-frequency domain, the sparsity property is more dominant to further reduce computational complexity.
This is also beneficial for ultra-massive MIMO channel in high mobility mmWave/THz band. The Doppler effect and high delay resolution can be well handled.
The accuracy and complexity of channel estimation can also be improved.

\subsection{Beam domain channel characterization and modeling}

Through reviewing existing literatures, we find that there lacks in-depth works regarding beam domain channel parameter estimation, channel characteristics analysis, and channel modeling of ultra-massive MIMO channel.
As they are fundamentals for the development of spatial-temporal transmission and random access technologies, it is working digging into: 1) the definition of effective beam domain parameters and parameter estimation methods that considering ultra-massive MIMO channel characteristics; 2) the manifestation of spatial domain channel characteristics in beam domain and exploration of other new beam domain channel characteristics with the combination of mmWave/THz in time-variant scenario; 3) the BDCMs that can accurately describe beam domain channel characteristics with reduced complexity than~GBSMs.

\subsection{Large dimensional computing theory}

A significant hinder lying behind massive MIMO system is the overwhelming computational burden.
With the increase of array elements, mathematical calculations in signal design, channel estimation, detection, resource allocation, etc., pose huge challenges to ultra-massive MIMO system.
Flexible and efficient computational framework for large dimensional computing theory is expected to support the development of ultra-massive MIMO system.
The striving directions may be how to utilize the inherent features of massive data and the advantages of AI algorithms.

\section{Conclusions}

In this work, recent advances of 6G ultra-massive MIMO techniques in both spatial and beam domains have been reviewed.
Therein, ultra-massive MIMO channel parameter estimation algorithms, channel characteristics, and channel models have been introduced as the foundations of further PHY technology design and evaluation.
Channel estimation, spatial multiplexing, and precoding techniques have been surveyed with the emphases on the transitions of data processing in the spatial domain into beam domain, thus to provide better tradeoff between accuracy and complexity.
However, more efforts need to be devoted to including new ultra-massive MIMO characteristics into consideration and clarifying the impacts of system setups on global performance.
Therefore, advanced PHY techniques can be proposed to reap the utmost benefits of ultra-massive MIMO communication system.

\begin{IEEEbiography}[{\includegraphics[width=1in,clip,keepaspectratio]{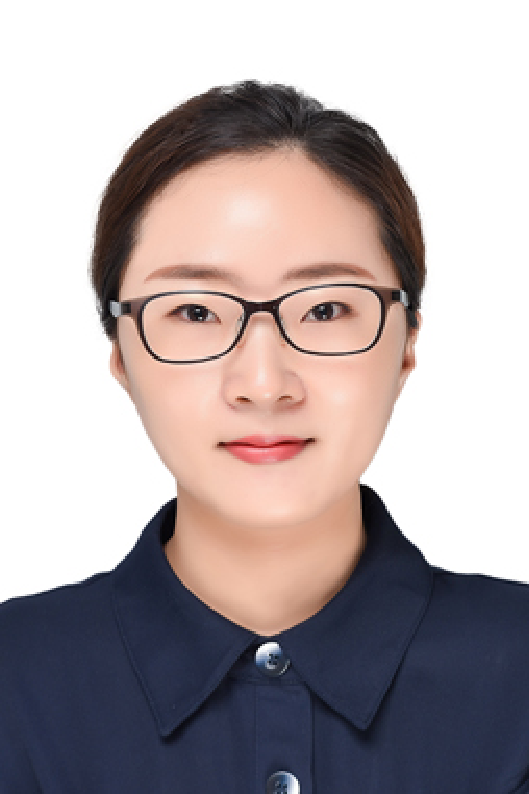}}]
{Rui Feng}
received the B.Sc. degree in Communication Engineering from Yantai University, China, in 2011, the M.Eng. degree in Signal and Information Processing from Yantai University, China, in 2014, and the Ph.D. degree in Communication and Information System from Shandong University, China, in 2018. From July 2018 to Sept. 2020, she was a lecture in Ludong University, China.
She is currently a Postdoctoral Research Associate in Purple Mountain Laboratories and Southeast University, China.
Her research interests include (ultra-) massive MIMO channel modeling theory and beam domain channel modeling.
\end{IEEEbiography}

\begin{IEEEbiography}[{\includegraphics[width=1in,clip,keepaspectratio]{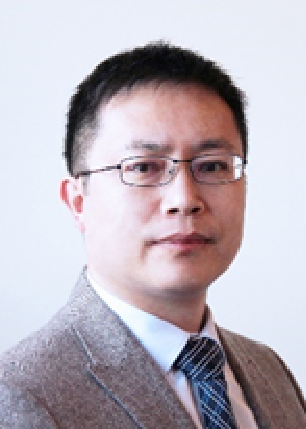}}]
{Cheng-Xiang Wang}
(F'17) received his Ph.D. degree from Aalborg University,	Denmark, in 2004. He joined Heriot-Watt University, Edinburgh, United Kingdom, in 2005 and became a professor in 2011. Since 2018, he has been with Southeast University, China, as a professor and is now the Executive Dean of the School of Information Science and Engineering. He is also a part-time professor with Purple Mountain Laboratories, China. His current research interests include wireless channel measurements and modeling, 6G wireless communication networks, and electromagnetic information theory. He is a Member of the Academia Europaea (The Academy of Europe), a Member of the European Academy of Sciences and Arts (EASA), a Fellow of the Royal Society of Edinburgh (FRSE), and IET, a Highly-Cited Researcher recognized by Clarivate Analytics in 2017--2020, and an Executive Editorial Committee Member of the IEEE TRANSACTIONS ON WIRELESS COMMUNICATIONS.
\end{IEEEbiography}

\begin{IEEEbiography}[{\includegraphics[width=1in,clip,keepaspectratio]{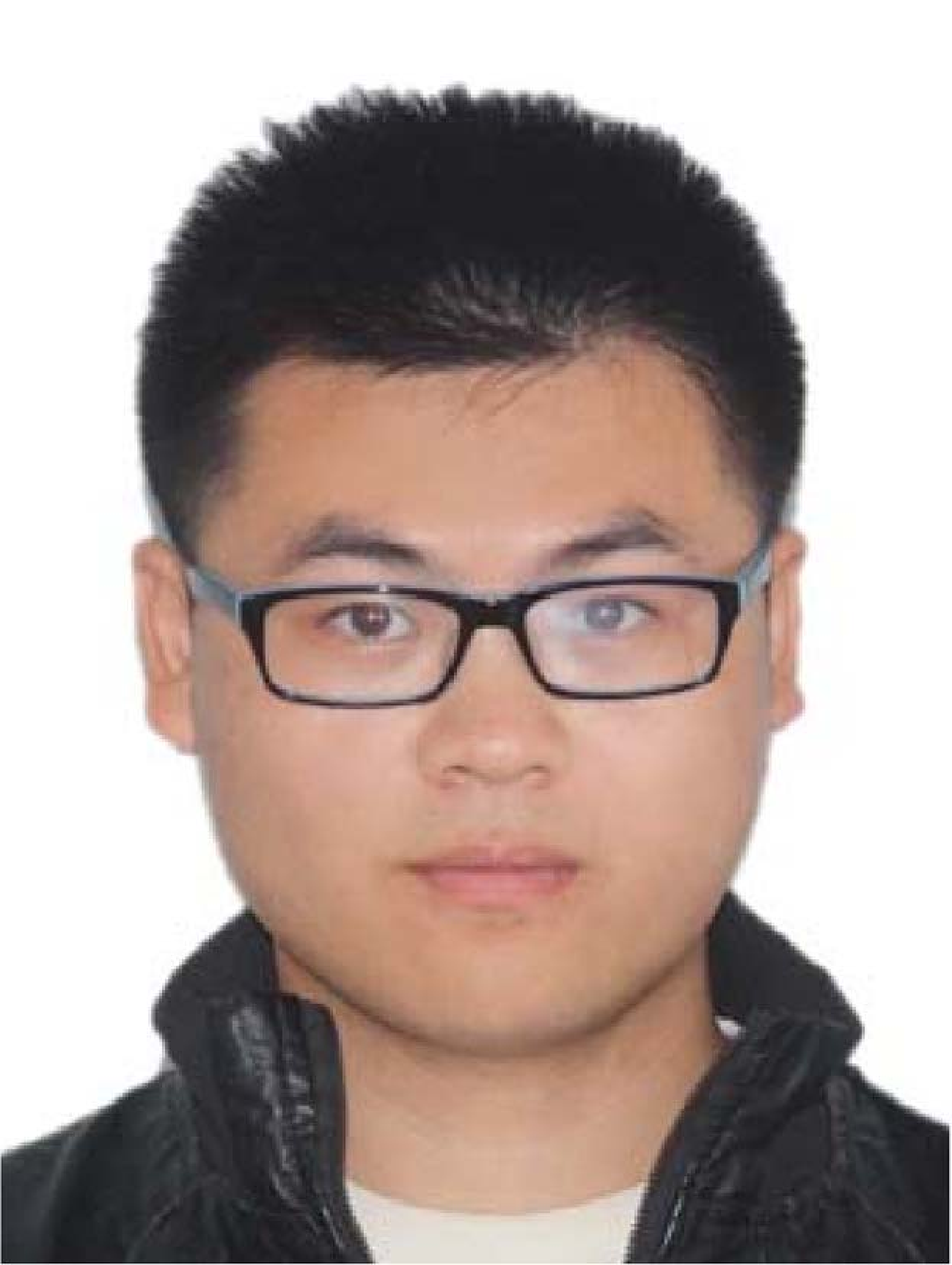}}]
{Jie Huang}
 received the B.E. degree in information engineering from Xidian University, China, in 2013, and the Ph.D. degree in communication and information systems from Shandong University, China, in 2018. From October 2018 to October 2020, he was a Research Associate with the National Mobile Communications Research Laboratory, Southeast University, China. From January 2019 to February 2020, he was a Research Associate with Durham University, U.K. He is currently an Associate Professor with the National Mobile Communications Research Laboratory, Southeast University, and also a Researcher with the Purple Mountain Laboratories, China. His research interests include millimeter wave, THz, massive MIMO, and intelligent reflecting surface channel measurements and modeling, wireless big data, and 6G wireless communications.
\end{IEEEbiography}

\begin{IEEEbiography}[{\includegraphics[width=1in,clip,keepaspectratio]{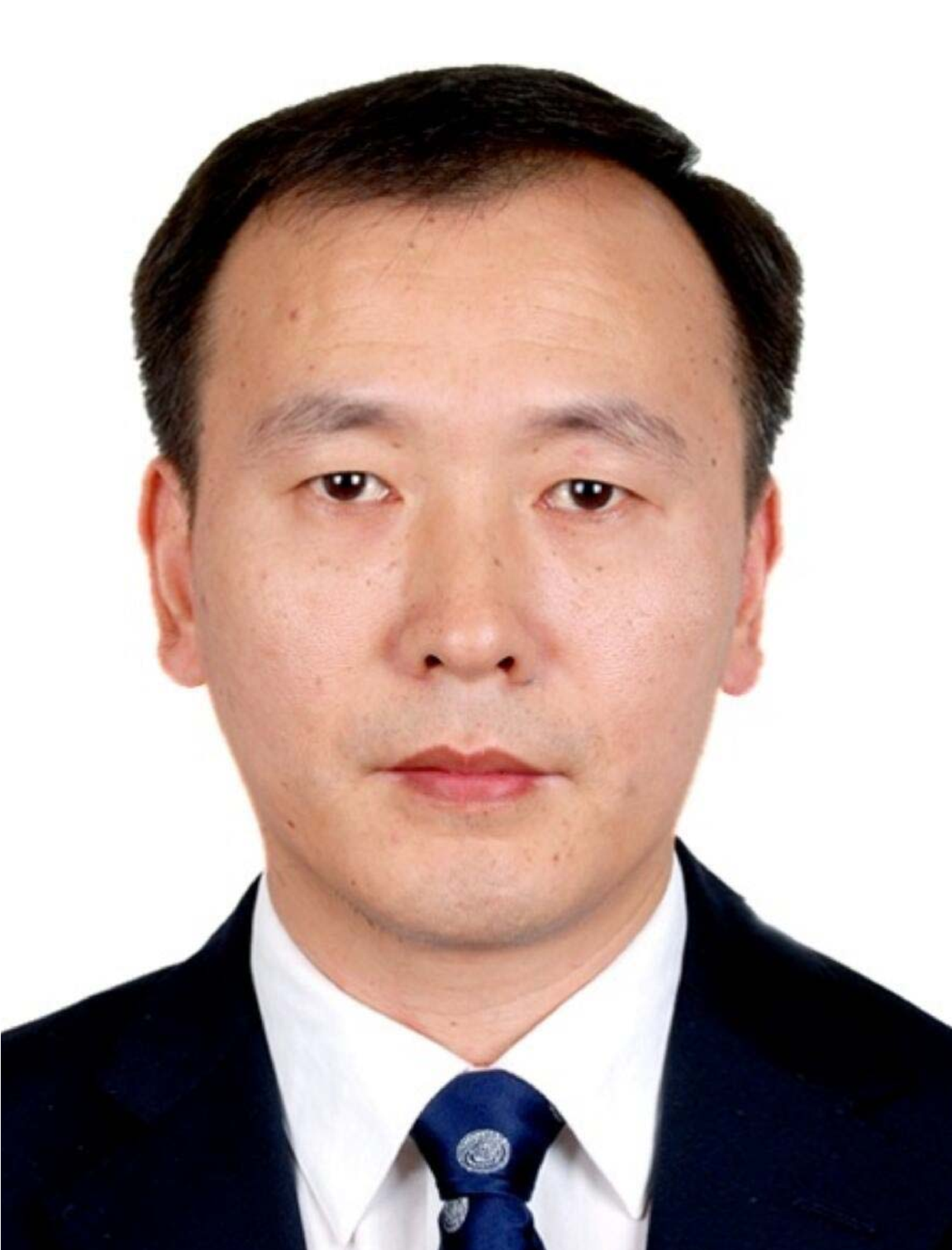}}]
{Xiqi Gao}
(F'15) received the Ph.D. degree in electrical engineering from Southeast University, Nanjing, China, in 1997. He joined the Department of Radio Engineering, Southeast University, in April 1992. Since May 2001, he has been a professor of information systems and communications. From September 1999 to August 2000, he was a visiting scholar at Massachusetts Institute of Technology, Cambridge, and Boston University, Boston, MA. From August 2007 to July 2008, he visited the Darmstadt University of Technology, Darmstadt, Germany, as a Humboldt scholar. His current research interests include broadband multicarrier communications, massive MIMO wireless communications, satellite communications, optical wireless communications, information theory and signal processing for wireless communications. From 2007 to 2012, he served as an Editor for the IEEE Transactions on Wireless Communications. From 2009 to 2013, he served as an Associate Editor for the IEEE Transactions on Signal Processing. From 2015 to 2017, he served as an Editor for the IEEE Transactions on Communications. Dr. Gao received the Science and Technology Awards of the State Education Ministry of China in 1998, 2006 and 2009, the National Technological Invention Award of China in 2011, the Science and Technology Award of Jiangsu Province of China in 2014, and the 2011 IEEE Communications Society Stephen O. Rice Prize Paper Award in the field of communications theory.
\end{IEEEbiography}

\end{document}